\documentclass{aa}  

\usepackage{graphicx}
\usepackage{txfonts}

\usepackage{gensymb}
\usepackage{multirow}

\usepackage{natbib,twoopt}
\usepackage[colorlinks,allcolors=blue]{hyperref} 

\usepackage{lscape}

\newcommand{\doce}{$^{12}$CO }
\newcommand{\docep}{$^{12}$CO}
\newcommand{\trece}{$^{13}$CO }
\newcommand{\trecep}{$^{13}$CO}
\newcommand{\dosuno}{\mbox{$J=$ 2\,--\,1}}
\newcommand{\unocero}{\mbox{$J=$ 1\,--\,0}}
\newcommand{\tresdos}{$J=3-2$}

\newcommand{\kms}{\,km\,s$^{-1}$ }
\newcommand{\kmsp}{\,km\,s$^{-1}$}
\newcommand{\ms}{\,M$_{\odot}$ }
\newcommand{\msp}{\,M$_{\odot}$}

\newcommand{\on}{89\,Her }

\newcommand{\fig}{Fig.\,}
\newcommand{\figs}{Figs.\,}
\newcommand{\tab}{Table\,}

\newcommand{\sect}{Sect.\,}

\newcommand{\secp}{\mbox{\rlap{.}$''$}}

\newcommand{\x}{\,$\times$\,}
\newcommand{\xd}[1]{10$^{#1}$}

\newcommand{\lsim}{\raisebox{-.4ex}{$\stackrel{\sf <}{\scriptstyle\sf \sim}$}}
\newcommand{\gsim}{\raisebox{-.4ex}{$\stackrel{\sf >}{\scriptstyle\sf \sim}$}}

\newcommand\sz{1}

\begin{document} 

   \title{The nebula around the binary post-AGB star 89\,Herculis\,\thanks{Based on observations with the 30\,m\,IRAM telescope and NOEMA. IRAM is supported by INSU/CNRS (France), MPG (Germany), and IGN (Spain).} \thanks{Final datacubes are available at the CDS via anonymous FTP
   to \url{cdsarc.u-strasbg.fr} (130.79.128.5) or via \url{http://cdsweb.u-strasbg.fr/cgi-bin/gcat?J/A+A/}}}
   \author{I. Gallardo Cava\,\inst{1}, J. Alcolea\,\inst{1}, V. Bujarrabal\,\inst{2}, M. Gómez-Garrido\,\inst{1,3}, and A. Castro-Carrizo\,\inst{4}}     

    \institute{Observatorio Astronómico Nacional (OAN-IGN), Alfonso XII 3, 28014, Madrid, Spain\\                 \email{i.gallardocava@oan.es}
    \and
    Observatorio Astronómico Nacional (OAN-IGN), Apartado 112, 28803, Alcalá de Henares, Madrid,     Spain   
    \and        
    Centro de Desarrollos Tecnológicos, Observatorio de Yebes (IGN), 19141, Yebes, Guadalajara,              Spain
    \and
    Institut de Radioastronomie Millimétrique, 300 rue de la Piscine, 38406 Saint-Martin-d'Hères, France        
}

\titlerunning{The nebula around the binary post-AGB star 89\,Herculis}
\authorrunning{Gallardo Cava, I. et al.}

   \date{}
   \date{Received 5 July 2022 / Accepted 5 January 2023}

 
  \abstract
   {There is a class of binary post-asymptotic giant branch (post-AGB) stars that exhibit remarkable near-infrared (NIR) excess. 
   These stars are surrounded by disks with Keplerian or quasi-Keplerian dynamics and outflows composed of gas escaping from the rotating disk.
   Depending on the dominance of these components, there are two subclasses of binary post-AGB stars: disk-dominated and outflow-dominated.
   }
   {We aim to properly study the hourglass-like structure that surrounds the Keplerian disk around 89\,Her.}
   {We present total-power on-the-fly maps of \doce and \trece \dosuno\ emission lines in 89\,Her. Previous studies are known to suffer from flux losses in the most extended components.
   We merge these total-power maps with previous NOEMA maps. 
   The resulting combined maps are expected to detect the whole nebula extent of the source.}
   {Our new combined maps contain the entirety of the detectable flux of the source and at the same time are of high spatial resolution thanks to the interferometric observations. 
   We find that the hourglass-like extended outflow around the rotating disk is larger and more massive than suggested by previous works. The total nebular mass of this very extended nebula is 1.8\x\xd{-2}\msp, of which $\sim$\,65\% comes from the outflow.
   The observational data and model results lead us to classify the envelope around 89\,Her as an outflow-dominated nebula, together with R\,Sct and IRAS\,19125+0343 (and very probably AI\,CMi, IRAS\,20056+1834, and IRAS\,18123+0511). 
The updated statistics on the masses of the two post-AGB main components reveal that there are two distinct subclasses of nebulae around binary post-AGB stars depending on which component is the dominant one. We speculate that the absence of an intermediate subclass of sources is due to the different initial conditions of the stellar system and not because both subclasses are in different stages of the post-AGB evolution.
   }
   {}

   

   \maketitle
%

\section{Introduction}
\label{introduccion}

Circumstellar envelopes around AGB stars (CSE-AGBs) are often found to be spherical and in slow expansion, while their descendants, that is pre-planetary and planetary nebulae (pPNe and PNe), present different morphologies: pPNe present strongly aspherical (often axisymmetrical) shapes resulting from the interaction of axial fast winds with the CSE formed in the AGB stage \citep[][]{ueta2000,sahai2007,castrocarrizo2010};
PNe tend to present (quasi-)spherical, axisymmetrical, or irregular shapes \citep[][]{sahai2011,stanghellini2016}.
This evolution is very rapid and takes place in a very short time, namely $\sim$\,1\,000\,years
\cite[see][]{balickfrank2002,vanwinckel2003}.
This spectacular transformation is thought to be due to magnetocentrifugal launching of outflows from rotating disks, which implies the presence of a stellar companion to provide the necessary amount of angular momentum \citep[see e.g.,][]{frankblackman2004, bujarrabal2016}.

There is a class of binary post-AGB stars (binary systems including a post-AGB star) that systematically present circumbinary disks with Keplerian dynamics.
These sources tend to present remarkable near-infrared (NIR) excess and narrow CO line profiles characteristic of rotating disks \citep[][and references therein]{vanwinckel2003,bujarrabal2013a}.
The IR data of these sources reveal the presence of highly processed dust grains, which implies that these disks must be stable structures \citep[][]{jura2003,sahai2011,gielen2011a}.

In the four sources in which the disk rotation has been well observed (the Red\,Rectangle, AC\,Herculis, IW\,Carinae, IRAS\,08544$-$4431), the disk contains most of the mass, but there is also gas in expansion, a disk wind that is extracted from the disk and represents \lsim\,15\% of the total nebular mass \citep[][]{bujarrabal2016, bujarrabal2017, bujarrabal2018, gallardocava2021}.
In contrast, there is a subclass of these binary post-AGB stars whose nebulae are dominated by the expanding component instead of the Keplerian disk: the outflow-dominated subclass \citep[see][]{gallardocava2021}.
In this latter class of sources, the nebula emission is quite different from that of the other well-studied cases (the disk-dominated nebulae), because their millimeter(mm)-wave interferometric data confirm the presence and dominance of an extended and expanding component that contains most of the detected nebular material.
This is the case for R\,Scuti and IRAS\,19125+0343, where $\sim$\,75\% of the total nebular mass corresponds to the extended and expanding component \citep[][]{gallardocava2021}. AI\,CMi, IRAS\,20056+1834, and IRAS\,18123+0511 also very probably belong to this subclass.
As we show in this paper, the last member of this (outflow-dominated) subclass is 89\,Herculis.

89\,Herculis is a binary post-AGB star with warm dust located in a stable structure and large dust grains formed and settled onto the midplane \citep[see][]{shenton1995,ruyter2006,hillen2014}.
This source has been thoroughly studied using single-dish observations. Studies reveal narrow CO line profiles similar to those of the Red\,Rectangle, but with prominent wings, suggesting a significant contribution of the extended component \citep[][]{bujarrabal2013a}. 
Observations at 1.3, 2, and 3\,mm reveal the presence of C$^{18}$O, C$^{17}$O, CS, SiS, and HCN and a very detailed analysis reveals that the nebula around 89\,Her is C-rich \citep[see][]{gallardocava2022}.
According to \citet{gallardocava2021}, the nebula around 89\,Her is composed of a Keplerian disk and an hourglass-shaped structure \citep[see also][]{alcoleabujarrabal1995,fong2006,bujarrabal2007}.
The mass of the nebula was found to be 1.4\x\xd{-2}\msp, of which 6.4\x\xd{-3}\ms corresponds to the Keplerian disk ($\sim$\,50\%). The findings of previous studies suggest that 89\,Her is the only source of our sample in which there is the same mass in the disk as in the disk wind.
However, the emission of the outflow was found to be underestimated because the NOEMA maps presented a significantly increased amount of filtered-out flux in the line wings in comparison to the single-dish 30\,m\,IRAM profile taken towards the center of the source.
Furthermore, the NOEMA maps show that the total extent of the outflow has a size comparable to the Half Power Beam Width (HPBW) of the 30\,m\,IRAM at the frequency of the CO \dosuno  transitions. 
Therefore, it is also possible that the lost flux problem of the interferometric observation is even more severe, as the 30\,m\,IRAM single-dish data may not contain all the flux emitted by the source, as NOEMA maps demonstrate that the source is larger than the beam. This implies that the mass derived for the outflow could be largely underestimated. 

To overcome this problem and properly derive the mass of the extended outflow component in a definitive way, here we present new single-dish total-power maps of 89\,Her. We merge these maps, which probe the full extent of 89\,Her, with the previous interferometric data, resulting in high-resolution maps of 89\,Her containing the total flux for all the nebular components detected in molecular gas.

\section{Observations and observational results} \label{observaciones}

As mentioned above, to solve the problem of the underestimation of the mass of the extended outflow component, it is necessary to construct maps of the source including the large spatial scales and not only those provided by the interferometric observations, which only probe structures of 5$''$ or smaller. To do this, we built new total-power single-dish maps of 92$''$ in diameter, and later combined these with the previously obtained interferometric maps already presented by \citet{gallardocava2021}.

\begin{figure}[h]
\center
\includegraphics[width=\sz\linewidth]{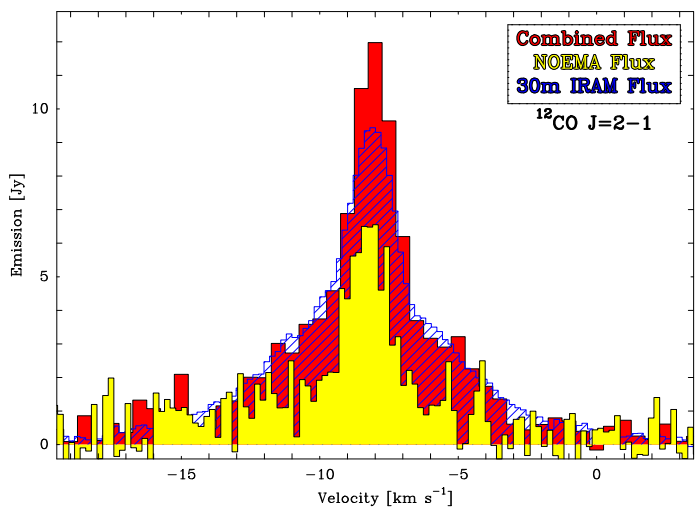} 
\caption{\small Histogram showing the flux comparison between the old NOEMA map (yellow), the old 30\,m\,IRAM single-pointing map (blue dashed), and NOEMA + the new total-power map (red), for \doce \dosuno.}
    \label{fig:flujos_doce}  
\end{figure}

\begin{figure}[h]
\center
\includegraphics[width=\sz\linewidth]{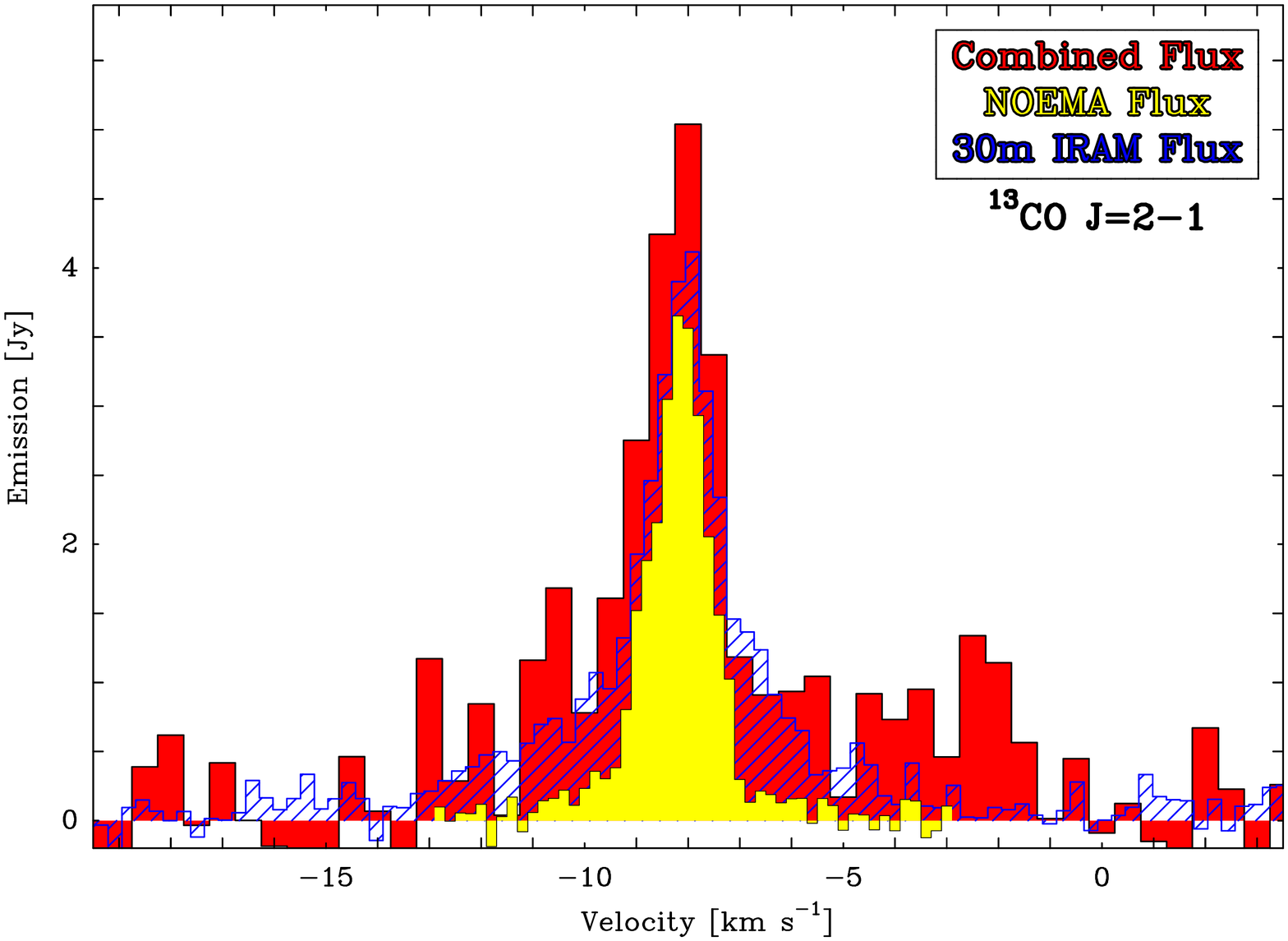}
\caption{\small Histogram showing the flux comparison between the old NOEMA map (yellow), the old 30\,m\,IRAM single-pointing map (blue dashed), and  NOEMA + the new total-power map (red), for \trece \dosuno.}
    \label{fig:flujos_trece}  
\end{figure}

\begin{figure*}[h]
\center
\includegraphics[width=\sz\linewidth]{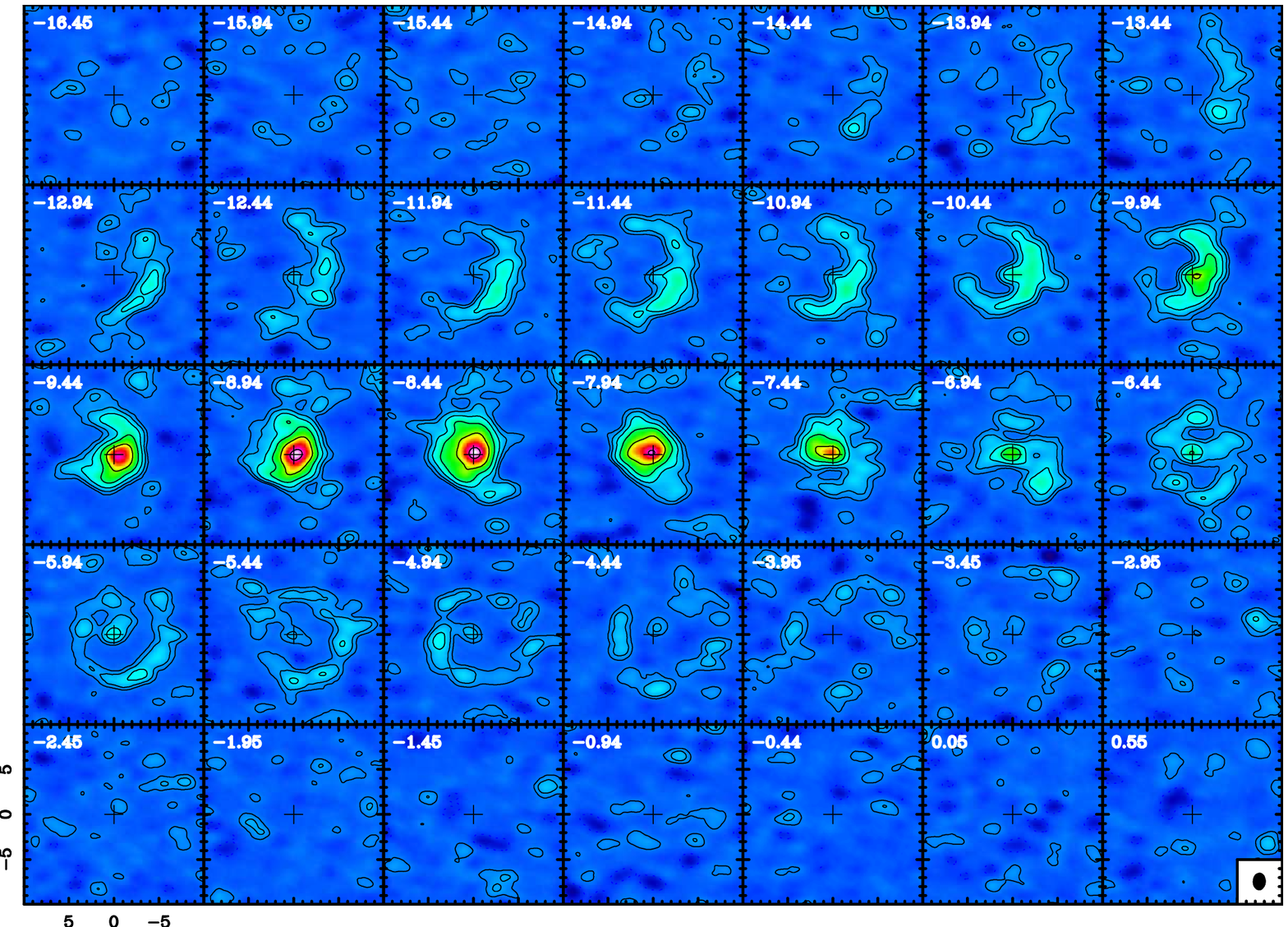}\caption{\small Merged (NOEMA + 30\,m) maps per velocity channel of \doce \dosuno\ emission from 89\,Her. 
The contours are $-$40, 40, 80, 160, 320, 640, and 1280\,mJy\,beam$^{-1}$, with a maximum emission peak of 1.5\,Jy\,beam$^{-1}$.
The resulting clean-beam has a HPBW of 1\secp{95} \x 1\secp{45}, the major axis oriented at PA = 92\degree. The LSR velocity is indicated in the upper right corner of each velocity-channel panel and the beam size is shown in the last panel. The FOV of each panel is $20''\times 20''$. X and Y units are offsets in arcseconds in the directions of  east and north, respectively.}
    \label{fig:mapas_doce}  
\end{figure*}

\begin{figure*}[h]
\center
\includegraphics[width=\sz\linewidth]{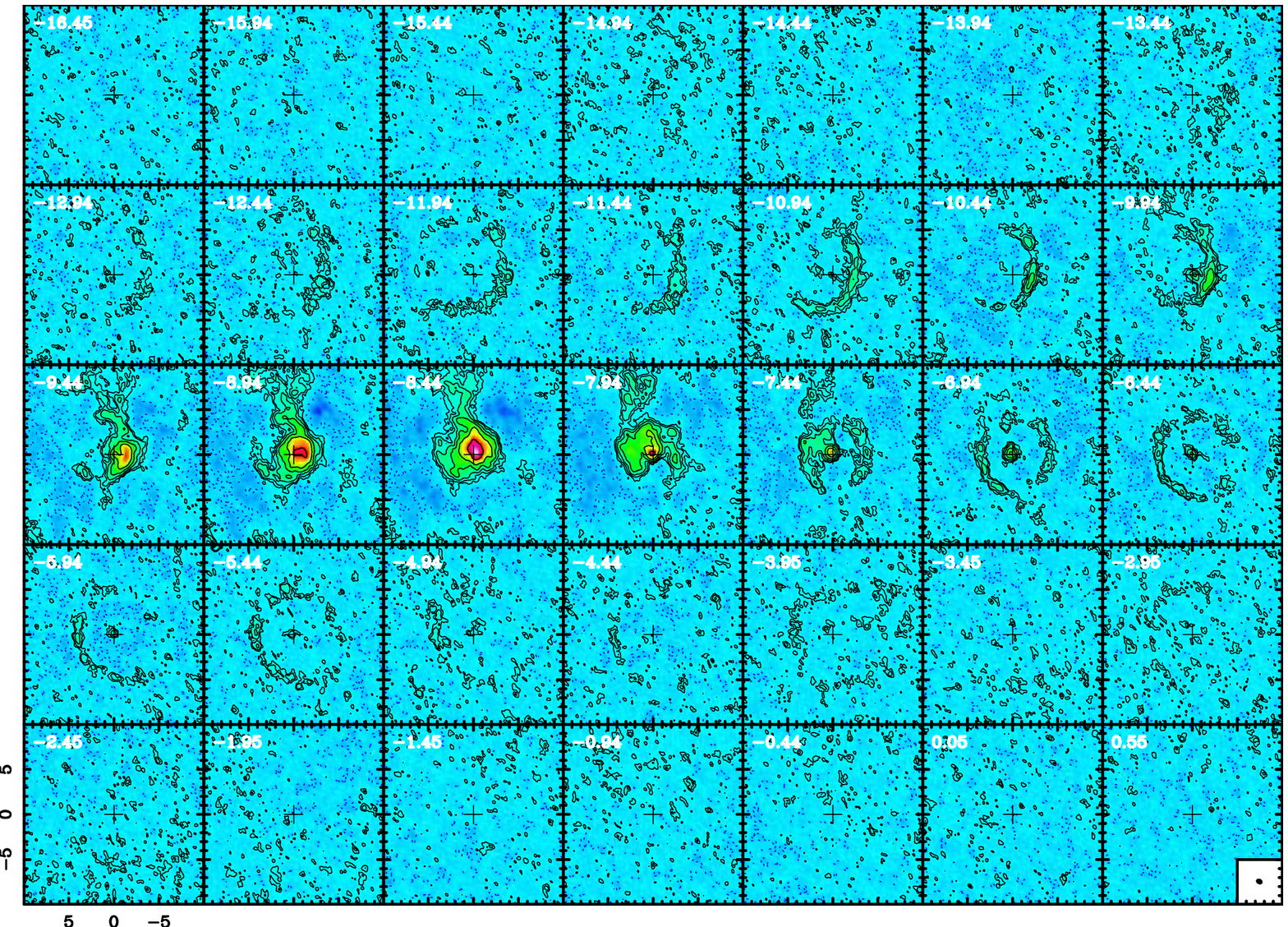}
\caption{\small Merged (NOEMA + 30\,m) maps  per velocity channel of \trece \dosuno\ emission from 89\,Her. 
The contours are $-$5, 5, 10, 20, 40, 80, and 160\,mJy\,beam$^{-1}$, with a maximum emission peak of 240\,mJy\,beam$^{-1}$.
The resulting clean-beam has a HPBW of 0\secp{76} \x 0\secp{59}, the major axis oriented at PA = 27\degree. The LSR velocity is indicated in the upper right corner of each velocity-channel panel and the beam size is shown in the last panel. The FOV of each panel is $20''\times 20''$. X and Y units
are offsets in arcseconds in the directions of  east and north, respectively.}
    \label{fig:mapas_trece}  
\end{figure*}

\subsection{Total-power data} \label{totalpower}

These new single-dish 89\,Her observations were performed using the 30\,m\,IRAM telescope at Pico Veleta (Granada, Spain).
The observations were carried out between June 2 and 8, 2021, for a total of 65 hours of telescope time (project 061-21). 
We focus our observations on the \doce and \trece \dosuno\ emission lines (230.5 and 220.4\,GHz), but we also simultaneously observed the \doce and \trece \unocero\ lines. However, the signal-to-noise ratio (S/N) attained for the \unocero\ lines did not result in their detection, and therefore these lines are not discussed in this paper.
The HPBW of the telescope is around 11\secp{3} at 230.5 and 220.4\,GHz.
During the observing time the weather conditions were good, with typical 225\,GHz zenital opacities of between 0.1  and 0.3.

We performed 92$''$\,$\times$\,92$''$ maps every 20\,min. The scanning of these maps was performed in six different directions, at paralactic angles (PAs) of 0\degree, 30\degree, 60\degree, 90\degree, 120\degree, and 150\degree, in order to minimize interleaving and weaving patterns in the resulting final averaged map. 
The observations were performed in on-the-fly (OTF) mode, with 23 parallel scans per map, and a separation between scans of 4$''$, a scanning velocity of 4$''$s$^{-1}$, and a dump rate of 1\,s, resulting in a 4$''$\,$\times$\,4$''$ grid (to be compared with the HPBW of 11\secp{3}, i.e., a factor 3 oversampling).
The observations were performed in on-off mode, observing the reference position (600$''$ away in the R.A. direction) every three scans ($\sim$\,70\,s). A calibration was performed at the beginning of each individual map using the chopper-wheel method, observing the sky and both hot and cold loads. Observations of NGC\,7027 were also performed to check the consistency of the intensity scale: As we did not find calibration changes larger that 20\%, no re-scale has been applied to the data. 

We connected the Fast Fourier Transform Spectrometer (FTS) units to the EMIR receiver with a spectral velocity resolution of 0.25\kms ($\sim$\,200\,kHz) per channel. We obtained spectra for vertical and horizontal linear polarization receivers; as no significant changes were found between the two polarizations, the two maps were averaged.
 
We applied the standard data-reduction procedure, which consists of the removal of baselines, calculating a polarization average, and  resampling data to the desired final spectral resolution. Individual OTF maps were inspected for consistency of the pointing: we found no significant differences in the position of the central peak, and so all maps were averaged without applying corrections. 

The OTF maps for the $J=2-1$ lines are shown in \figs\ref{fig:mapas_doce_tp} and \ref{fig:mapas_trece_tp}.
As we can see in \fig\ref{fig:mapas_doce_tp}, the extent of the nebula is restricted to a region of 20$''$ in diameter. This is fully compatible with the extent of the nebula in the interferometric maps considering the resolution of the OTF data. This relatively large extent also justifies the flux loss in the NOEMA data. Moreover, this extent is larger than the HPBW of the 30\,m\,IRAM, which explains the larger total flux in the new OTF observations in comparison with the old single-pointing 30\,m data. 
On the contrary, this size is smaller than the HPBW of the primary beam of the individual NOEMA antennas at 1.3\,mm ($\sim$\,22$''$). Nevertheless, the primary beam shape attenuation is taken into account, because it is corrected before the process of merging with the total-power data, following the standard procedure of the GILDAS software\footnote{GILDAS is a software package designed to reduce and analyze mainly mm observations from single-dish and  interferometric telescopes. It is developed and maintained by IRAM, LAOG/Université de Grenoble, LAB/Observatoire de Bordeaux, and LERMA/Observatoire de Paris. See \url{https://www.iram.fr/IRAMFR/GILDAS}}.

\subsection{Interferometric data}

Interferometric observations of the \doce and \trece \dosuno\ rotational transitions were carried out toward 89\,Her with the IRAM NOEMA interferometer at Plateau de Bure (Grenoble, France).
\doce \dosuno\ observations were performed under project name P05E, while \trece \dosuno\ observations were performed under project names X073 and W14BT. 
These maps were published by \citet{gallardocava2021}, where the complete technical description can be found.

\subsection{Combined maps}

In this work, we present NOEMA maps of $^{12}$CO and $^{13}$CO $J=2-1$ emission lines, which include short-spacing pseudovisibilities obtained from the total-power maps with the 30\,m\,IRAM through OTF mode. Therefore, our combined maps contain all detectable flux because (a) they recover the lost flux of the extended component filtered out by the interferometer and
(b) include large areas that single-dish single-pointing observations cannot detect because of the limitations of the beam of the telescope; see \figs\ref{fig:flujos_doce} and \ref{fig:flujos_trece}.

To include the large scales probed by the OTF maps into the interferometric data, these total-power maps must first be converted into pseudo-visibility data cubes with exactly the same velocity configuration (same number of channels, velocity spacing, and velocity value for the reference channel) as the NOEMA data. We did this by resampling the OTF maps into a spectral resolution of 0.5\kms and adopting a LSR velocity of $-$7.94\kms for the central channel. Then, data cubes with a pixel size of 2$''$\,$\times$\,2$''$ were built from the OTF scans, which are Fourier transformed for obtaining the corresponding pseudo-visibilities to be merged with the interferometric visibilities from the corresponding NOEMA observations. Before the merging, these interferometric data are corrected for the NOEMA primary beam attenuation following the standard procedure of GILDAS.

\begin{figure}[t]
\center
\includegraphics[width=\sz\linewidth]{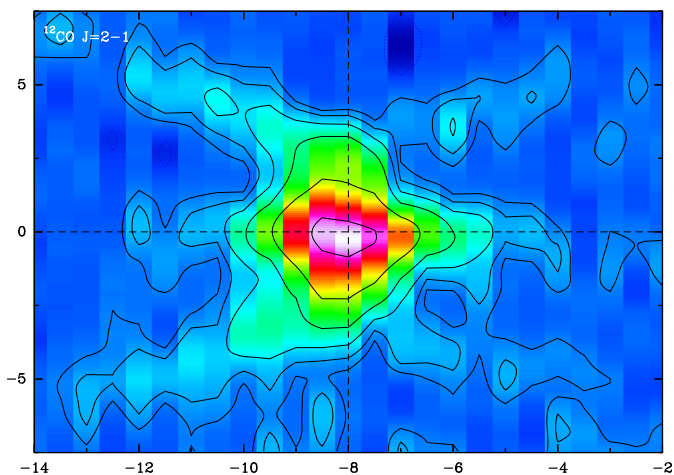}
\caption{\small Position--velocity diagram from our merged maps of \doce \dosuno\ in 89\,Her along the direction PA = 150\degree, corresponding to the nebula equator. The contours are: $-$40, 40, 80, 160, 320, 640, and 1280\,mJy\,beam$^{-1}$, with a maximum emission peak of 1.45\,Jy\,beam$^{-1}$. The dashed lines show the central position and systemic velocity.}
    \label{fig:pv_doce_ec}  
\end{figure}

\begin{figure}[t]
\center
\includegraphics[width=\sz\linewidth]{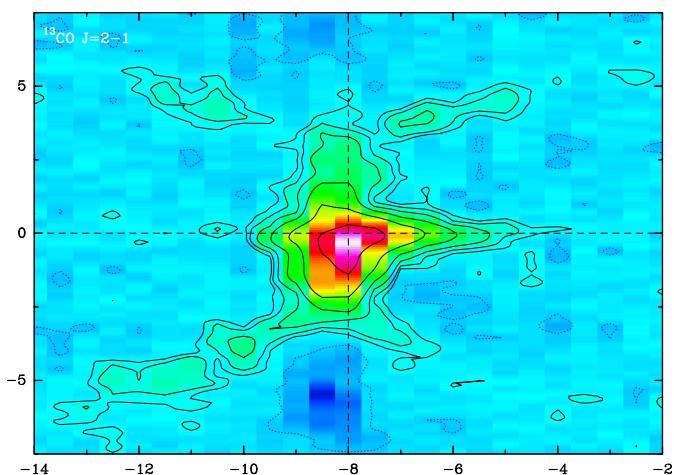}
\caption{\small Position--velocity diagram from our merged maps of \trece \dosuno\ in 89\,Her along the direction PA = 150\degree, corresponding to the nebula equator. The contours are $-$5, 5, 10, 20, 40, 80, and 160\,mJy\,beam$^{-1}$, with a maximum emission peak of 240\,mJy\,beam$^{-1}$. The dashed lines show the central position and systemic velocity.}
    \label{fig:pv_trece_ec}  
\end{figure}

After the merging of the $uv$ data sets, we verified the compatibility of the calibration of the two instruments, 30\,m\,IRAM and NOEMA. The new maps were then produced by first mapping and then cleaning the merged uv data sets.
For the \doce \dosuno\ maps, we used robust weighting and a $uv$ tapper of 200\,m, resulting in a HPBW synthetic clean beam of 1\secp{95} \x 1\secp{45}.
For the case of \trece \dosuno, as the interferometric data were of better quality, we used robust weighting and no tappering.
This resulted in a HPBW synthetic clean beam of 0\secp{76} \x 0\secp{59}.
The resulting new merged maps for \doce \dosuno\ and \trece \dosuno\ are presented in \figs\ref{fig:mapas_doce} and \ref{fig:mapas_trece}, respectively.
We verified that the new maps are compatible with the old interferometric data  and that the total flux is the same as in the new OTF maps, that is, no flux is missing in the new data.
We therefore conclude that our new combined maps contain all the detectable flux, because they are not affected by any interferometric losses and include the full extent of the source. This new total flux is larger than that obtained from the interferometric observation alone and that provided by single-dish, single-pointing observations; see \figs\ref{fig:flujos_doce} and \ref{fig:flujos_trece}.
In addition, position--velocity (PV) diagrams for PA $= 150\degree$ are shown in \figs\ref{fig:pv_doce_ec} and \ref{fig:pv_trece_ec}.

These velocity maps and equatorial PV diagrams reveal an intense central clump and an expanding component.
On the one hand, the central clump corresponds to an unresolved rotating disk very probably with Keplerian dynamics.
On the other hand, we also see an expanding component that very nicely shows hourglass-like features.
Following the same reasoning as in our previous work \citep[][]{gallardocava2021}, we find that PV diagrams along a PA of 150\degree\ are the most efficient at revealing the presence of the rotating gas (see \figs\ref{fig:pv_doce_ec} and \ref{fig:pv_trece_ec}). Therefore, the PV diagrams with PA = 150\degree suggest that a rotating disk with moderate velocity dispersion is responsible for the compact and central clump in  89\,Her.
Thanks to our new combined maps, we confirm the presence of asymmetrical velocities in both maps, because more emission is present at positive velocities than at negative ones. This phenomenon is a result of self-absorption by cold gas in expansion located in front of the rotating component
and is often found to be present in the disk-containing sources that we have observed in the past, such as the Red\,Rectangle or R\,Sct, a disk- and an outflow-dominated source, respectively  \citep[see][]{bujarrabal2016, gallardocava2021}.

Our combined maps, with all possible detected flux, show an hourglass outflow whose symmetry axis is almost aligned with our line of sight. These maps also reveal that the size of this outflowing component is larger than we found it to be in a previous study, both in the height of the hourglass and in the width of its expanding walls. This recovered flux has a relevant impact on the mass of the outflow that surrounds the disk (see \sect\ref{modelo}).

Apart from the rotating disk and the very large hourglass-like expanding component, we do not detect any other structures (such as a halo).
We verified this by performing radial averages of the emission in the OTF maps. The resulting intensity-versus-offset profiles obtained for both \doce and \trece\  \dosuno\ maps are compatible with the 89\,Her nebula consisting only in the two structures mentioned above.
Therefore, we conclude that the molecular envelope of 89\,Her comprises an hourglass-shaped outflow and a rotating disk with Keplerian dynamics.

We note that the S/N is higher in the new combined maps, which allows detection of the outflow at slightly larger expansion velocities. The only new feature worth mentioning is the tentative detection of a spiral-like pattern seen at some receding velocities ($-$5.95\kmsp) in the \doce\dosuno\ maps.
If real, this structure cannot be explained as a result of the orbital motions in the binary system, because this would imply a period much larger (1\,800\,years for an expansion velocity of $\sim$\,10\kms and a distance of 1\,000\,pc) than that derived from the radial velocity curve of the primary \citep[289.1\,days; see][]{oomen2018}.
This structure could instead be due to the effect of a precessing jet launched from the compact companion \citep[see][]{raga2009,velazquez2011}. However, this should be investigated more thoroughly using more sensitive maps.
This is similar to the case of the pattern seen in the Red\,Rectangle, which also suggests a periodicity much larger than the orbital period in the system \citep[see][]{cohen2004,waelkens1996}.

\section{Model fitting: Mass of the nebula} \label{modelo}

To derive the physical parameters of the nebula from the new data presented here, we adopted a nebular model based on a rotating disk with Keplerian dynamics surrounded by an extended and expanding hourglass-shaped structure. We use the code described in \citet{gallardocava2021}, where a complete description can be found. As in the previous work, we adopt a distance of 1\,000\,pc.
We assume LTE populations for the \dosuno\ CO lines, which is a reasonable assumption for low-$J$ rotational levels, simplifies the calculations, and provides a more comprehensible interpretation of the fitting parameters.
We adopt the same relative abundance values as in  \citet{gallardocava2021}: X(\trecep) = 2\x\xd{-5} and an abundance ratio of X(\docep)/X(\trecep) = 10. These abundance values are usually found in nebulae around binary post-AGB stars.

We assume the presence of a rotating disk in the innermost region of the nebula and a large and wide hourglass-like extended and expanding component.
We confirm the inclination of the nebular symmetry axis with respect to the line of sight, together with PA = 150\degree\ for the equatorial direction.
We also corroborate the self-absorption effects at low negative expansion velocities.

The main parameters of our new best model can be seen in \tab\ref{89hermodel} and can be described as follows. As in the first work, we find the same reliable results for density and temperature laws with high slope values.
The density of the rotating disk is assumed to vary with the distance to the binary system following a potential law, $n \propto r^{-2}$, with a value of 2.0\x\xd{7}\,cm$^{-3}$ at 2.5\x\xd{15}\,cm.
The temperature varies as $T \propto r^{-2.5}$, with a temperature of 425\,K at 5\x\xd{15}\,cm (the disk radius).
Moreover, we assume Keplerian rotation in the disk, with 1.5\kms at 10$^{16}$\,cm, which is compatible with a central total stellar mass of 1.7\msp.

\begin{figure}[t]
\center
\includegraphics[width=\sz\linewidth]{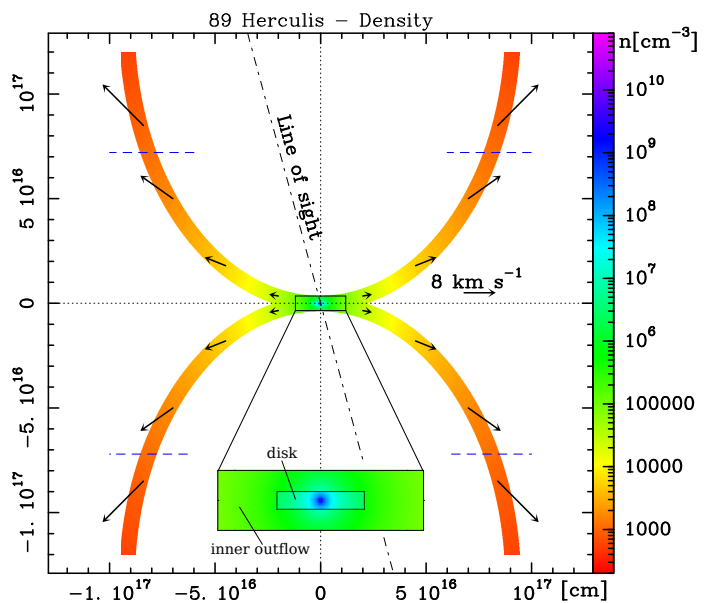}
\caption{\small Structure and distribution of the density of our best-fit model for the disk and outflow of 89\,Her. 
The lower inset shows a zoom into the innermost region of the outflow together with the rotating disk, with density values of $\sim$\,10$^{6}$ and $\geq$\,10$^{7}$\,cm$^{-3}$, respectively.
The rest of the outflow presents density values $\leq$\,10$^{5}$\,cm$^{-3}$.
The expansion velocity is represented with arrows.
The four horizontal blue dashed lines mark the former extent of the nebula in our previous model \citep[see][]{gallardocava2021}.}
    \label{fig:89her_dens_otf}  
\end{figure}

In the case of the outflow/disk wind, we assume a radial velocity with a modulus that increases linearly with the distance to the center, and we find moderate velocities of $\sim$\,10\kmsp.
We assume an hourglass 67\% larger with walls 10\% wider compared to the previous work (see \fig\ref{fig:89her_dens_otf}).
Therefore, the total height of the hourglass is 2.4\x\xd{17}\,cm and its walls are $\sim$\,8\x\xd{15}\,cm wide.
The density law varies as $n \propto r^{-2}$, with values in between 10$^{5}$\,cm$^{-3}$ in the zones closest to the rotating disk and $\leq$\,10$^{3}$\,cm$^{-3}$ in the most external regions of the outflow.
The temperature of the outflow varies as $T \propto r^{-0.2}$, with values $\leq$\,10\,K in the external regions of the outflow. We therefore find a very extended outflow composed of cold gas, as in the case of R\,Sct or IRAS\,19125+0343, which are similar objects \citep[see][]{gallardocava2021}.

We show the predictions from our best model in our synthetic velocity maps (\figs\ref{fig:mapas_doce_modelo} and \ref{fig:mapas_trece_modelo}) and synthetic PV diagrams (\figs\ref{fig:pv_doce_ec_modelo} and \ref{fig:pv_trece_ec_modelo}) and our new nebula model for 89\,Her in \fig\ref{fig:89her_dens_otf}.
We also checked that our model satisfactorily reproduces previous interferometric maps of the \doce \unocero\ emission line \citep[data taken from][whose maps also show no flux loss]{fong2006}.
The model reproduces the observational data and yields a total mass for the nebula of 1.8\x\xd{-2}\msp, of which 1.2\x\xd{-2}\ms corresponds to the outflow and 6.4\x\xd{-3}\ms to the rotating disk. This means that 89\,Her presents a Keplerian disk surrounded by a very extended and expanding outflow that represents $\sim$\,65\% of the total mass of the nebula. 

Some of the nebula properties are not accurately determined because of the relatively low resolution of the interferometric data, such as the height of the Keplerian disk. On the contrary, the outflow structure is well determined, because the hourglass-like shape is clear in our maps. See \citet{gallardocava2021}, for further details on the uncertainties of our model.

\begin{table}[t]
 \caption{Physical conditions in the molecular Keplerian disk and expanding hourglass-like component of \on derived from our best-fit model of the CO data.}
\tiny
\vspace{-5mm}
\begin{center}

\begin{tabular*}{\linewidth}{llll}
\hline \hline
 \\[-2ex]
Parameter   & Disk   & Outflow  & Outflow (old) \\ \hline
 \\[-2ex]
 
\multirow{2}{*}{Radius\,[cm]} & \multirow{2}{*}{$5.0\times 10^{15}$} &  $R_{\text{max}}=9.5 \times 10^{16}$ &  $8.6 \times 10^{16}$\\
 \vspace{1mm}
 & & $W_{\text{o}}=0.8$\x\xd{16} & 0.7\x\xd{16}\\
  \vspace{1mm}
Height\,[cm] & $1.0\times 10^{15}$  & $12\times 10^{16}$ & $7.2\times 10^{16}$ \\
\multirow{2}{*}{Density\,[cm$^{-3}$]} & $n_{0}=2.0\times 10^{7}$ &  $n_{0}=1.5\times 10^{3}$ & $n_{0}=3.0\times 10^{3}$\\
 \vspace{1mm}
 & $\kappa_{\text{n}}=2.0$   & $\kappa_{\text{n}}=2.0$ &  $\kappa_{\text{n}}=1.8$ \\
\multirow{2}{*}{Temperature\,[K]} & $T_{0}=75$ & $T_{0}=5.5$ & $T_{0}=10$ \\
 \vspace{1mm}
 & $\kappa_{\text{T}}=2.5$ & $\kappa_{\text{T}}=0.2$  &  $\kappa_{\text{T}}=0$ \\
  \vspace{1mm}
 Rot.\,Vel.\,[km\,s$^{-1}$] & 1.5  & $-$ \\
  \vspace{1mm}
Exp.\,Vel.\,[km\,s$^{-1}$] & $-$ & 1.2  \\  
 \vspace{1mm}
X(\docep) & 2.0\x10$^{-4}$ & 2.0\x10$^{-4}$ \\
 \vspace{1mm}
\docep\,/\,\trecep & 10 & 10 \\
 \vspace{1mm}
Inclination\,[\degree] & & 15 \\

Position angle\,[\degree] & & 60 \\

\hline
 
\end{tabular*}

\end{center}
\small
\vspace{-1mm}
\textbf{Notes.} Parameters and their values used in our best-fit model for the disk and outflow. We note that we also quote the values that have changed from our previous (old) work.
See \citet{gallardocava2021} for a complete description of the model and its parameters.
\label{89hermodel}
\end{table}

\section{Disk-mass ratio in binary post-AGB stars}

In our previous work, we classified 89\,Her as an intermediate-subclass nebula, that is, intermediate between the disk- and the
outflow-dominated subclasses,  because both the disk and its outflow were found to present very similar masses. However, a significant amount of  flux was found to be filtered out in those maps (see \sect \ref{introduccion}).
According to our new merged maps and the new model, the pPN around 89\,Her is dominated by its disk wind, because $\sim$\,65\% of the total mass is located in the very large hourglass-like expanding component.

After this update on the masses of both outflow and disk in 89\,Her, this source is more in line with what is found in those classified as outflow-dominated. In fact, if we only consider sources for which these masses are relatively well determined (see \tab\ref{masas} and \fig\ref{fig:histo_outflow}), it seems that there is not an intermediate case in between the disk- and the outflow-dominated subclass. However, before drawing conclusions, it is better to revise any problem in the estimation of the masses of the two components, such as the impact of and hypothetical loss of flux (as in the case of 89\,Her, which we review immediately above), the impact of other nebular components ---in particular the neutral and ionized atomic gas---, or the contribution of the layers where molecules have been photo-dissociated by the interstellar UV field.

We reiterate that the main goal of this work is to study the circumbinary disk and the outflow or disk wind that is escaping from the rotating component during the post-AGB phase. 
Other nebular components, such as collimated high-velocity stellar winds (launched by the companion) or larger haloes (formed during the previous AGB phase and the detection of which could be quite difficult), are not part of this work.
In any case, we discuss the relevance of these structures and their (possible) contribution to the total nebular mass.

\subsection{The effects of the interferometric flux losses}

Seven sources have been thoroughly studied through interferometric observations: AC\,Her, the Red\,Rectangle, IRAS\,08544$-$4431, IW\,Car, 89\,Her, IRAS\,19125+0343, and R\,Sct.
Here, we discuss the flux loss present in the maps used to estimate the molecular mass in each source of our sample and update the values of the mass of the outflow and consequently the disk-to-outflow mass ratio. In the following, we assume that the filtered-out flux is due to the outflow, as no other large-scale structures have been detected in these kinds of objects.

\subsubsection{AC\,Herculis}

The \doce\dosuno\ mm-wave interferometric observations of this source do not present a significant amount of filtered-out flux. In addition, its angular size is around 2$''$, which is too small compared with the beam size. Therefore, we can confidently discard the presence of a very extended and undetected component.
According to this, the total molecular nebular mass is 8.3\x\xd{-4}\ms and the derived disk-mass ratio of AC\,Her of $\sim$\,95\% is well determined. See \citet{gallardocava2021} for further details.

\subsubsection{The Red\,Rectangle} \label{rr}

This source is the best studied of our sample. The nebular mass was derived from model fitting of maps of lines \doce \tresdos\ and $J=6-5$, \trece \tresdos, C$^{17}$O $J=6-5$, and H$^{13}$CN $J=4-3$. The total nebular mass is 1.4\x\xd{-2}\msp, of which $\sim$\,9\% corresponds to the outflow that surrounds the Keplerian disk. 
\citet{bujarrabal2016} mentioned that they found 25\% flux loss in the line wings. If we assume that this filtered-out flux corresponds to the outflow, the contribution of the outflow to the total molecular mass of the nebula is up to 11\%. Therefore, the Red\,Rectangle remains as a disk-dominated source.

\subsubsection{IRAS\,08544$-$4431}

The nebular mass of this source is 2.0\x\xd{-2}\ms \citep[][]{bujarrabal2018}. According to the authors, the mass of the outflow is 10\%. The \doce\tresdos\ maps present a small fraction of flux loss ($<$\,20\%), and the \trece\tresdos\ maps, with a less extended brightness distribution, do not show a significant amount of filtered-out flux. 
Assuming that this flux loss is due to the outflow, its contribution to the total molecular mass will be $\leq$\,12\%.
Additionally, the angular size of this source is around 4$''$, which is considerably smaller than the beam size.

\subsubsection{IW\,Carinae}

This source, of 4$''$ in angular size, presents a total nebular mass of 4\x\xd{-3}\msp, of which 11\% is located in the expanding component \citep[see][for details]{bujarrabal2017}.
The authors find a small fraction of filtered-out flux (25\%) in the \doce\tresdos\ maps, which could be higher in the line wings.
If this flux loss corresponds to the expanding component, it mass contribution is $\leq$\,14\% of the total nebular mass. Additionally, \citet{bujarrabal2013a}, via single-dish observations, found very narrow CO line profiles characteristic of rotating disk and weak wings that correspond to the outflow. This source would still be a disk-dominated source even in in the least favorable flux-loss scenario.

\subsubsection{89\,Herculis}

In this work, we present interferometric maps merged with large-scale single-dish maps. These combined maps do not present filtered-out flux (see \sect\ref{observaciones}).
Our new combined maps and model allow us to estimate the mass of the nebula that surrounds 89\,Her. We find a total molecular mass of 1.8\x\xd{-2}\msp, in which the hourglass-shaped outflow represents 65\% of the mass (see \sect\ref{modelo}).

\subsubsection{IRAS\,19125+0343}

This nebula presents a total mass of 1.1\x\xd{-2}\ms and 71\% corresponds to the outflow \citep[see][for details]{gallardocava2021}. The interferometric visibilities were merged with zero-spacing data obtained with the 30\,m\,IRAM telescope. Therefore, there is no filtered-out flux in the \doce\dosuno\ final maps, and so this source is clearly an outflow-dominated source.

\subsubsection{R\,Scuti}

The nebular mass is 3.2\x\xd{-2}\msp, of which 73\% is located in the extended outflow that surrounds the Keplerian disk \citep[see][for complete description]{gallardocava2021}. The \doce\dosuno\ maps do not present filtered-out flux, because they were merged with short-spacing pseudo-visibilities obtained from on-the-fly maps with the 30\,m\,IRAM telescope to compensate for the extended component filtered out by the interferometer, as this source presents a large angular size. This case is similar to that of 89\,Her.

\begin{table}[t]
 \caption{Sample of binary post-AGB stars for which the disk-mass ratio has been analyzed.}
\tiny
\vspace{-5mm}
\begin{center}

\begin{tabular*}{\columnwidth}{lcccl}
\hline \hline
 \\[-2ex]
\multirow{2}{*}{Source}   & $M_{\text{neb}}$   & Outflow  & $d$  & \multirow{2}{*}{Subclass} \\ 
                          &   [M$_{\odot}$]    & [\%]     & [pc] &         \\ \hline
 \\[-2ex]
 
 AC\,Herculis       &  8.3\x\xd{-4}  &  5  & 1100 & Disk-dominated     \\
 Red\,Rectangle     &  1.4\x\xd{-2}  &  11  &  710 & Disk-dominated     \\
 IRAS\,08544$-$4431 &  2.0\x\xd{-2}  & $\leq$12  & 1100 & Disk-dominated     \\
 IW\,Carinae        &  4.0\x\xd{-3}  & $\leq$14  & 1000 & Disk-dominated     \\ 
 89\,Herculis       &  1.8\x\xd{-2}  & 65  & 1000 & Outflow-dominated  \\
 IRAS\,18123+0511$^{*}$   &  4.7\x\xd{-2}  & 70  & 3500 & Outflow-dominated \\
 IRAS\,19125+0343   &  1.1\x\xd{-2}  & 71  & 1500 & Outflow-dominated  \\
 R\,Scuti           &  3.2\x\xd{-2}  & 73  & 1000 & Outflow-dominated  \\
 AI\,Canis\,Minoris$^{*}$ &  1.9\x\xd{-2}  & 75  & 1500 & Outflow-dominated  \\ 
 IRAS\,20056+1834$^{*}$   &  1.0\x\xd{-1}  & 75  & 3000 & Outflow-dominated  \\ 

\hline
 
\end{tabular*}

\end{center}
\small
\vspace{-1mm}
\textbf{Notes.} Sources marked with an asterisk have been studied only via single-dish observations and the values of their masses and outflow-mass ratios can present higher uncertainties.
\label{masas}
\end{table}

\subsection{Atomic mass}

Some sources of our sample have been studied in the \textsc{C\,ii} line (157.14\,$\mu$m). 
The flux of this line can be used to measure the mass of the low-excitation atomic component in pPNe, because this transition is useful for estimating the total mass of photodissociation regions (PDRs).
PDRs almost perfectly coincide with the cold atomic gas regions, because they are very clearly delimited zones in between the molecular and the \textsc{H\,ii} regions. In addition, most of our sources show relatively low stellar temperatures, of namely $<$\,10\,000\,K, and so \textsc{H\,ii} regions are not expected in them. We note that in \citet{gallardocava2022}, no radio recombination lines were detected in these spectral surveys.
\citet{castrocarrizo2001} and \citet{fong2001} found that the PDR masses for AC\,Her, the Red\,Rectangle, 89\,Her, and R\,Sct (rescaled to our distances; see \tab\ref{masas}) are $<$\,10$^{-2}$, $<$\,10$^{-2}$, $<$\,2\x\xd{-3}, and  $<$\,3\x\xd{-2}\msp.
\citet{bujarrabal2016} detected \textsc{C\,ii} in the Red\,Rectangle and, using the method described in \citet{castrocarrizo2001}, the mass of the PDR derived from the measured flux would represent $<$\,10$^{-3}$\ms which would mostly be located in the rotating disk.

We conclude that there are no indications of a significant contribution from PDRs or \textsc{H\,ii} regions to the masses of the disk-containing nebulae around binary post-AGB stars.
Therefore, there is no need to consider these small contributions in the disk-mass ratio analysis.

\subsection{Post-AGB stellar winds} \label{stellarwinds}

Systematic high-resolution radial velocity studies reveal that stellar jets or winds are a common phenomenon in binary post-AGB stars \citep[see][and references therein]{bollen2021}. A collimated low-density and high-velocity jet launched by the companion often operates during the late AGB and early post-AGB stages \citep[see][and their Fig.\,1]{bollen2022}.
We note that this is a nebular component not studied in this work (focused on rotating disks and disk winds). However, we argue that the mass of this component is small.

This kind of stellar wind is present in the best studied source of our sample, the Red\,Rectangle, and can be clearly seen in the optical image \citep[][]{cohen2004}. At first sight, the H$\alpha$ emission of the famous X-shaped structure seems to dominate the whole nebula.
Nevertheless, according to the high velocities and mass-loss found in this stellar wind \citep[$\sim$\,200\kms and 10$^{-6}$\,M$_{\odot}$\,a$^{-1}$, respectively\footnote{We follow the recommendations for units of the IAU Style Manual \citep[][]{wilkins1990, wilkins1995}. Therefore, we use the term annus, abbreviated to ``a'', for year.}, see][]{thomas2013,bollen2019}, we deduce that this jet is \lsim\,1\,000 years old and its mass is \lsim\,10$^{-3}$\msp.
Although this stellar wind is not part of our study, its mass is small compared with the mass derived in this work for the rotating disk, and is slightly lower than that of the molecular outflow.
In addition, we highlight that the ``X''-shaped structure seen in \citet{cohen2004} only shows the gas that is closest to the axis, while CO interferometric observations reveal that the optical X-shaped structure represents the inner layer of the outflowing biconical component \citep[][]{bujarrabal2016}.
Moreover, \citet{Menshchikov2002}, from a detailed model of this nebula in the polar direction, found that the gas and dust density values sharply decay from 2$''$. This is roughly consistent with the strong decrease in brightness seen in \citet{cohen2004} at $\pm$\,2$''$, as the optical brightness of the outermost regions of the Red\,Rectangle ($\pm$\,10$''$) is more than 50 times weaker than that at $\pm$\,2$''$.

In view of what is stated in \sect\ref{rr}, the rotating disk dominates the Red\,Rectangle nebula. As for the expanding gas, we clearly differentiate between an outflow composed of low-velocity expanding gas escaping from the disk, and these collimated high-velocity stellar winds that do not contribute notably to the total nebular mass (and their study is not part of this work).
Therefore, we prefer not to take this contribution into account, as it does not significantly affect our results.
We expect a similar or even more favorable situation for all the sources in our study. For instance, the size of 89\,Her via CO data seems to be compatible with that in the optical or IR image.

\subsection{Possible presence of haloes}

It is thought that \lsim\,1\ms of mass-loss from the AGB phase is necessary to reach the post-AGB stage. However, the masses detected in this study are small compared to what is predicted from theoretical works. One might wonder whether or not there is a large component surrounding the material that we detect.

The origin of these large quasi-spherical fossil shells requires successive ejecta of stellar material when the star reached the tip of the AGB phase \citep[][]{guerreromanchado1999}. 
This structure can be difficult to detect and its mass can vary between 10$^{-4}$ and $\sim$\,0.1\msp; and its contribution to the total mass may not be negligible
\citep[][]{chu1991,bujarrabal2001,guerrero2003,cox2012,vandesteene2015}.
Models of dust emission from our sources and standard pre-planetary nebulae show significant differences, with no hint in our disk-containing objects of extended shells, which are characteristic of standard pPNe; see e.g., \citet{gezer2015, hrivnak1989} (as in general detected in CO lines). 
In particular, from far-infrared (FIR) data in \citet{gezer2015}, in particular the emission excess at 20\,$-$\,100\,$\mu$m in their Fig.\,2, one finds that the mass of the undetected outer-shell in 89\,Her must be more than 100 times lower than that of the well-known pPN HD\,161796. We reiterate the fact that the \textit{Gaia} distance of HD\,161796 is $\sim$\,2\,kpc and that its nebula is supposed to be placed at 1\,$-$\,3\x\xd{17}cm, just beyond the nebula we detect in 89\,Her. As the mass of the HD\,161796 nebula is around 0.1\ms \citep[from CO data in][and again accounting for the new \textit{Gaia} distance]{bujarrabal2001}, we derive a lower limit for the mass for any very outer shell around 89\,Her, \lsim\,10$^{-3}$\msp, which is a small value compared with the mass values derived in this paper for the rotating disk and disk wind.

The objects of our sample could present the same well-known missing mass problem as the other evolved sources \citep[see e.g.,][and references therein]{santandergarcia2021}. This missing mass could reside in a hard-to-detect halo. 
Future ultrasensitive observations and theoretical works will be needed to answer this question.
In any case, these haloes, which by definition arise from mass loss during the AGB stage, are unrelated to the post-AGB structures studied in this work (rotating disks and outflows).

\subsection{Contribution of outer layers of the disk wind} \label{capasexternas}

We remind the reader that, according to our previous discussion, the contribution of very wide components is not significant, in particular in the case of 89\,Her \citep[see also][]{hrivnak1989,gezer2015}.
In our analysis, the contribution of very outer layers, where CO can be photodissociated by interstellar UV, has not been included. This is a very general problem in the study of circumstellar envelopes, because these cool, dissociated layers are very difficult to detect and analyze. We think that the contribution of such outer haloes is likely negligible. General studies of FIR dust emission \citep[see e.g.,][]{gezer2015, vickers2015}, which is known to be useful for identifying very outer layers beyond the CO photodissociation radius, do not reveal the presence of sizable very extended components in binary post-AGB stars. 

Additionally, \citet{Menshchikov2002} modeled  the dust emission from the Red\,Rectangle in regions far from the poles in great detail, including an extended component in expansion, and found that the density decreases very steeply from a radius of 1\,000\,to 2\,000\,AU, which is very similar to what was derived from CO maps \citep[][]{bujarrabal2016}. 
Seemingly, the contribution of outer photodissociated shells to the total mass budget should be negligible in our sources.

Moreover, from the CO maps, we know that the kinetic time of the CO outflows in binary post-AGB stars is of a few thousand years.  According to the evolutionary models, this timescale is comparable to the total expected time spent by a star in its transition from the AGB to the PN phase \citep[see][]{bloeckerschoenberner1990,bertolamimarcelo2016}. High-mass Keplerian disks are a typical post-AGB phenomenon, being very rare or absent in AGB stars, and the ejection of the observed outflow from the disk must also be a post-AGB phenomenon.

In conclusion, our results so far suggest that we cannot expect any greatly extended outflows in the objects we have  studied. Nevertheless, we recognize that the analysis of these difficult-to-detect layers is uncertain.
Even in the hypothetical case where the mass of outer and undetected layers of the outflow is comparable to that of the detected ones (i.e. a factor of 2), the contribution of the disk wind to the total nebular mass would always be $<$\,25\% in the case of the disk-dominated sources. On the contrary, this outflow mass ratio would be \gsim\,80\% in the case of the outflow-dominated ones.
Therefore, even in this hypothetical scenario, our main conclusions of the disk-to-mass ratio fraction would not change significantly.

\subsection{Bimodal distribution of the wind contribution}

\begin{figure}[t]
\center
\includegraphics[width=\sz\linewidth]{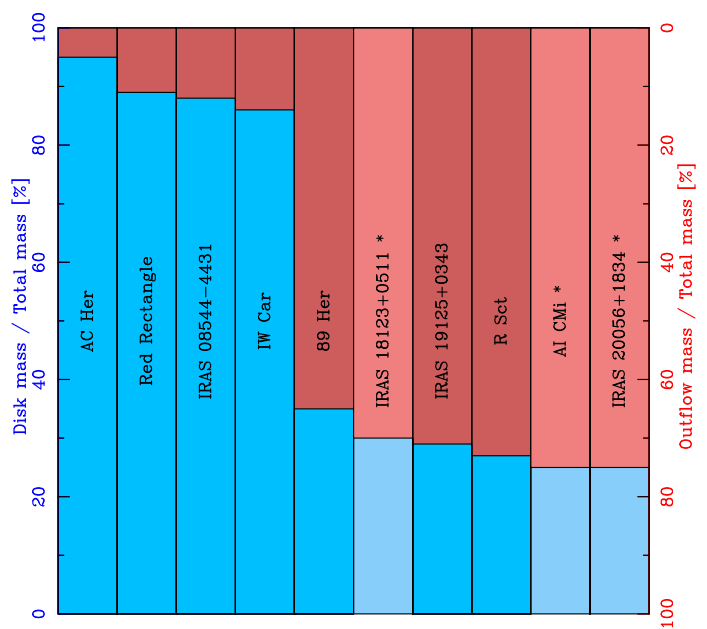}
\caption{\small Histogram showing the mass percentage of the disk (blue) and outflow (red) components for our sample of nebulae around binary post-AGB stars.
Two subclasses are clearly differentiated: the disk- and outflow-dominated sources.
Sources marked with an asterisk (and lighter colors) have been studied only through single-dish observations, and their values are less certain.}
    \label{fig:histo_outflow}  
\end{figure}

Following this very detailed analysis of each object, we update the total mass and contribution of the outflow (i.e., the disk wind) for the sample. These updated values can be seen in \tab\ref{masas} and \fig\ref{fig:histo_outflow}.
In particular, the figure shows that there is a clear bimodal distribution of the contribution of the disk wind to the total nebular mass around binary post-AGB stars. There are two types of sources; disk-dominated and outflow-dominated sources.
We note that the segregation between the two subclasses is very clear. The disk-dominated sources always present disk-mass percentages above 15\%, whereas in the outflow-dominated ones, this value is always below 40\%.
More precisely, the disk-dominated sources present an outflow contribution of 10\,$\pm$\,5\%, while in the outflow-dominated ones this figure is 70\,$\pm$\,10\%.

We note that there are no intermediate sources in between the two subclasses; see \tab\ref{masas} and \fig\ref{fig:histo_outflow}.
The absence of intermediate sources indicates that the two subclasses of sources are probably not related via an evolutionary link; otherwise one would expect to find intermediate sources because of the similar  outflow and post-AGB lifetimes (\sect\ref{capasexternas}). Therefore, we suggest that the existence of these two subclasses is related to the initial stellar properties and configuration of the binary system, resulting in the ability to launch a more or a less massive wind from the rotating disk. Of course, such reasoning would have to be revised if the post-AGB life scales of our sources are demonstrated to be much longer than for standard pPNe, at least by an order of magnitude, which cannot be excluded; to explain the bimodal distribution, we would then have to assume that the mass loss takes place in a critically small fraction of the post-AGB evolution.

\section{Conclusions} \label{conclusiones}

The new combined maps of 89\,Her with no flux loss lead us to study the very large hourglass-like extended component that surrounds the Keplerian disk. We find a total nebular mass of 1.8\x\xd{-2}\msp, of which 65\% is located in the outflow. According to these new mass estimations, 89\,Her has been re-classified as an outflow-dominated source.

After a very detailed analysis of each object of our sample of nebulae around binary post-AGB stars, we clearly find two subclasses: the disk- and outflow-dominated sources.
\fig\ref{fig:histo_outflow} shows a histogram that indicates the mass percentage of the rotating disk and outflow components for our sample, showing that  85\%\,$-$\,95\% 
 of the mass in disk-dominated sources is located in the rotating component, while the rotating component of the outflow-dominated sources only contains 25\%\,$-$\,35\% of the total mass (HD\,52961 and IRAS\,19157$-$0247 have not been considered in this work because they were classified as intermediate sources with a high level of uncertainty, and they could belong to either subclass).

The nebulae around binary post-AGB stars present a bimodal distribution (see \fig\ref{fig:histo_outflow}). The existence of these two very different subclasses does not support an evolutionary relationship between them, as the timescale of the post-AGB evolution is believed to be very short, and comparable to the time required to form these disk winds. On the contrary, the existence of these two subclasses could be due to a different initial configuration of the binary system or different initial masses. We propose that the outflow-dominated sources could result from the presence of a very efficient mass loss from the disk in certain objects, the cause of which remains unknown. It is possible that the outflow would be particularly conspicuous in objects that have spent a relatively long time in the post-AGB phase and have then had time to form very extended, 
high-mass outflows.

\begin{acknowledgements}
We are grateful to the anonymous referee for the relevant recommendations and comments on the paper.
This work is based on observations of IRAM telescopes. IRAM is supported by INSU/CNRS (France), MPG (Germany), and IGN (Spain).
This work is part of the AxiN and EVENTs\,/\,NEBULAE\,WEB research programs supported by Spanish AEI grants AYA\,2016-78994-P and PID2019-105203GB-C21, respectively.
IGC acknowledges Spanish MICIN the funding support of BES2017-080616.
\end{acknowledgements}

\bibliographystyle{aa}
\bibliography{referencias}

\appendix

\section{\doce and \trece \dosuno\ total-power maps}

\begin{figure*}[!h]
\centering
\includegraphics[width=\sz\linewidth]{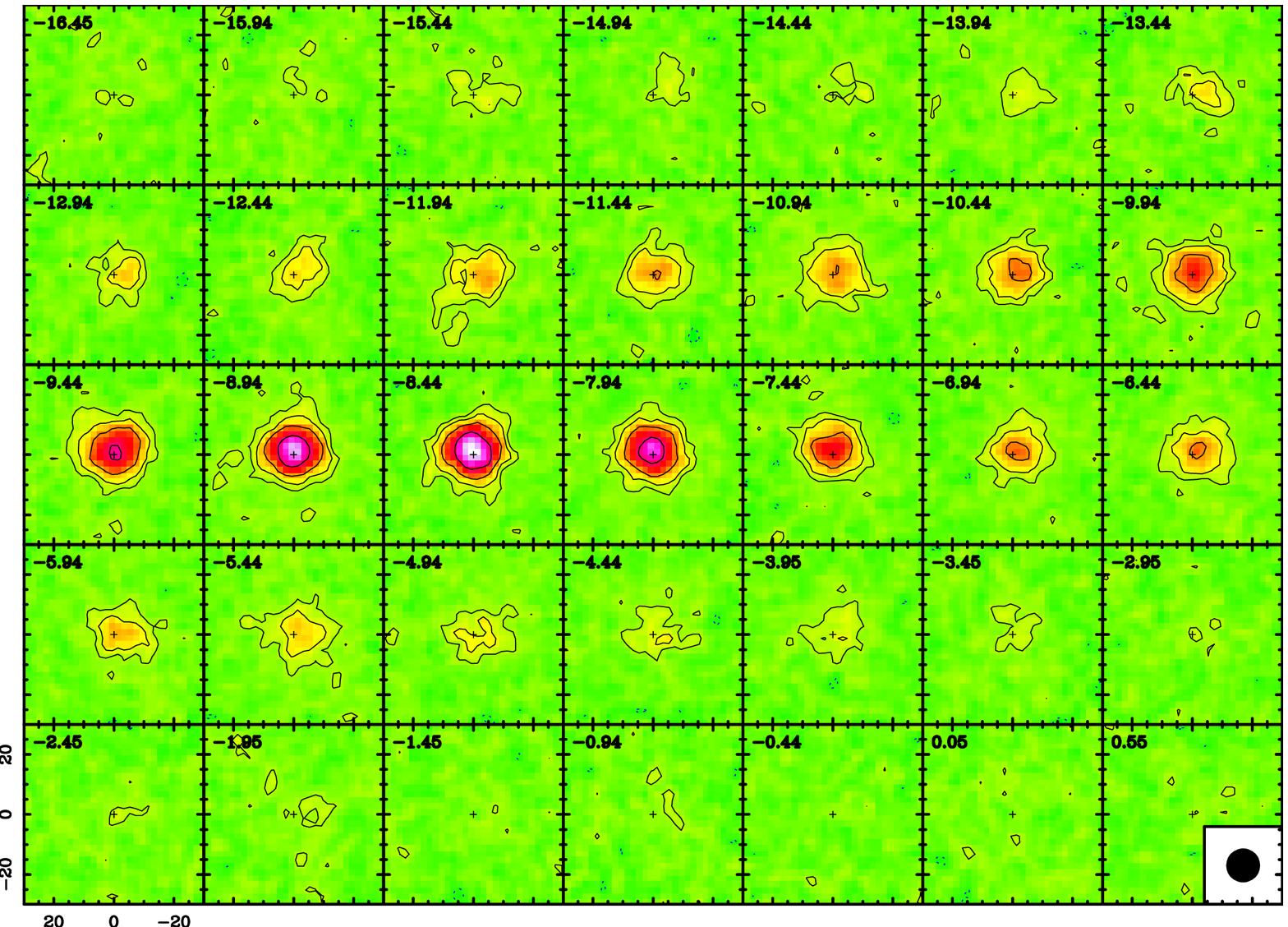}
\caption{\small Total-power maps per velocity channel of \doce \dosuno\ emission from 89\,Her. 
The contours are $-$0.1, 0.1, 0.2, 0.4, and 0.8\,K in main-beam scale.
The beam size (HPBW) is 11\secp{25} \x 11\secp{25}. The LSR velocity is indicated in the upper right corner of each velocity-channel panel and the beam size is shown in the last panel. The FOV of each pannel is $60''\times 60''$.}
    \label{fig:mapas_doce_tp}  
\end{figure*}

\begin{figure*}[!htpb]
\centering
\includegraphics[width=\sz\linewidth]{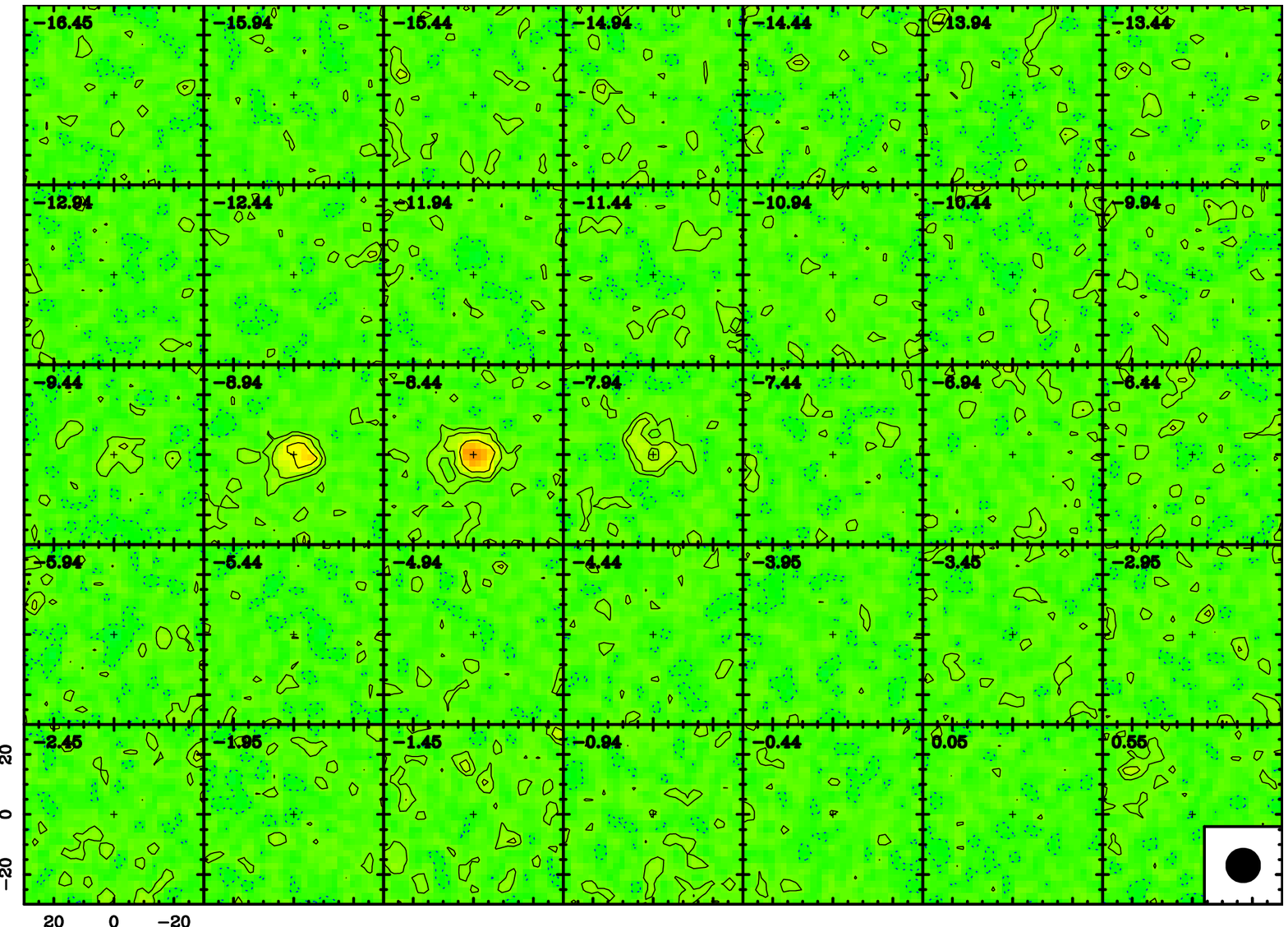}
\caption{\small Total-power maps per velocity channel of \trece \dosuno\ emission from 89\,Her. 
The contours are $-$0.08, 0.08, 0.16, and 0.32\,K in main-beam scale.
The beam size (HPBW) is 11\secp{77} \x 11\secp{77}. The LSR velocity is indicated in the upper right corner of each velocity-channel panel and the beam size is shown in the last panel. The FOV of each pannel is $60''\times 60''$.}
    \label{fig:mapas_trece_tp}  
\end{figure*}

Here we present total-power maps for 89\,Her of the \doce and \trece \dosuno\ emission lines; see \figs\ref{fig:mapas_doce_tp} and \ref{fig:mapas_trece_tp}. These observations were performed in OTF mode using the 30\,m\,IRAM telescope.
We detect emission at angular distances of around 20$''$ in diameter. Our new total-power maps present expansion velocities, which reveal a larger expanding component. See \sect\ref{totalpower} for a  complete description.

\clearpage
\section{Model results}

\begin{figure*}[!h]
\centering
\includegraphics[width=\sz\linewidth]{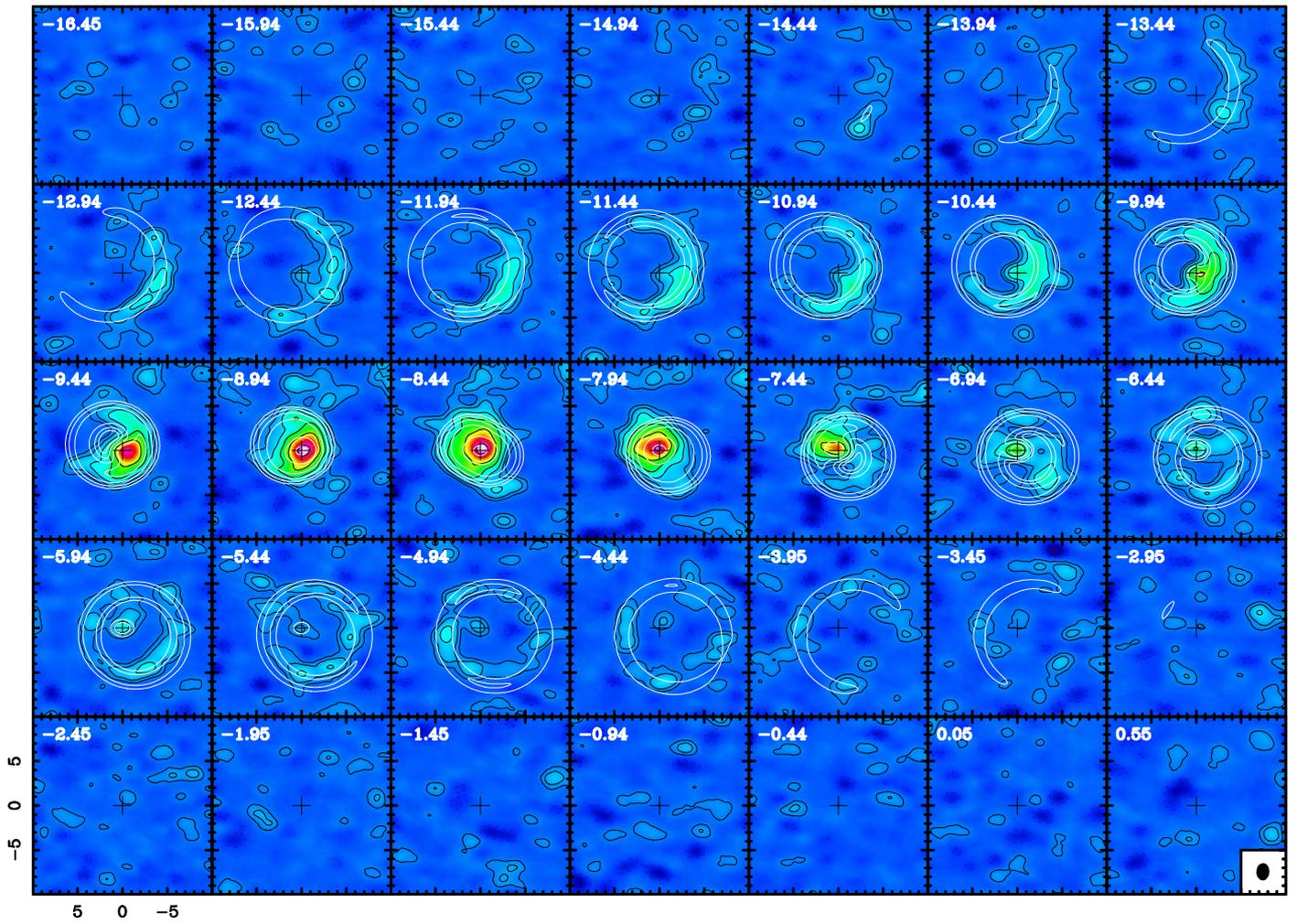}
\caption{\small Synthetic maps predicted by the model of the \doce \dosuno\ line emission for the nebula around 89\,Her, which is shown in white. The maps are superimposed on the observational ones. The contours for both the observational data and model are: $-$40 (observational data only), 40, 80, 160, 320, 640, and 1280\,mJy\,beam$^{-1}$.}
    \label{fig:mapas_doce_modelo}  
\end{figure*}

\begin{figure*}[!h]
\centering
\includegraphics[width=\sz\linewidth]{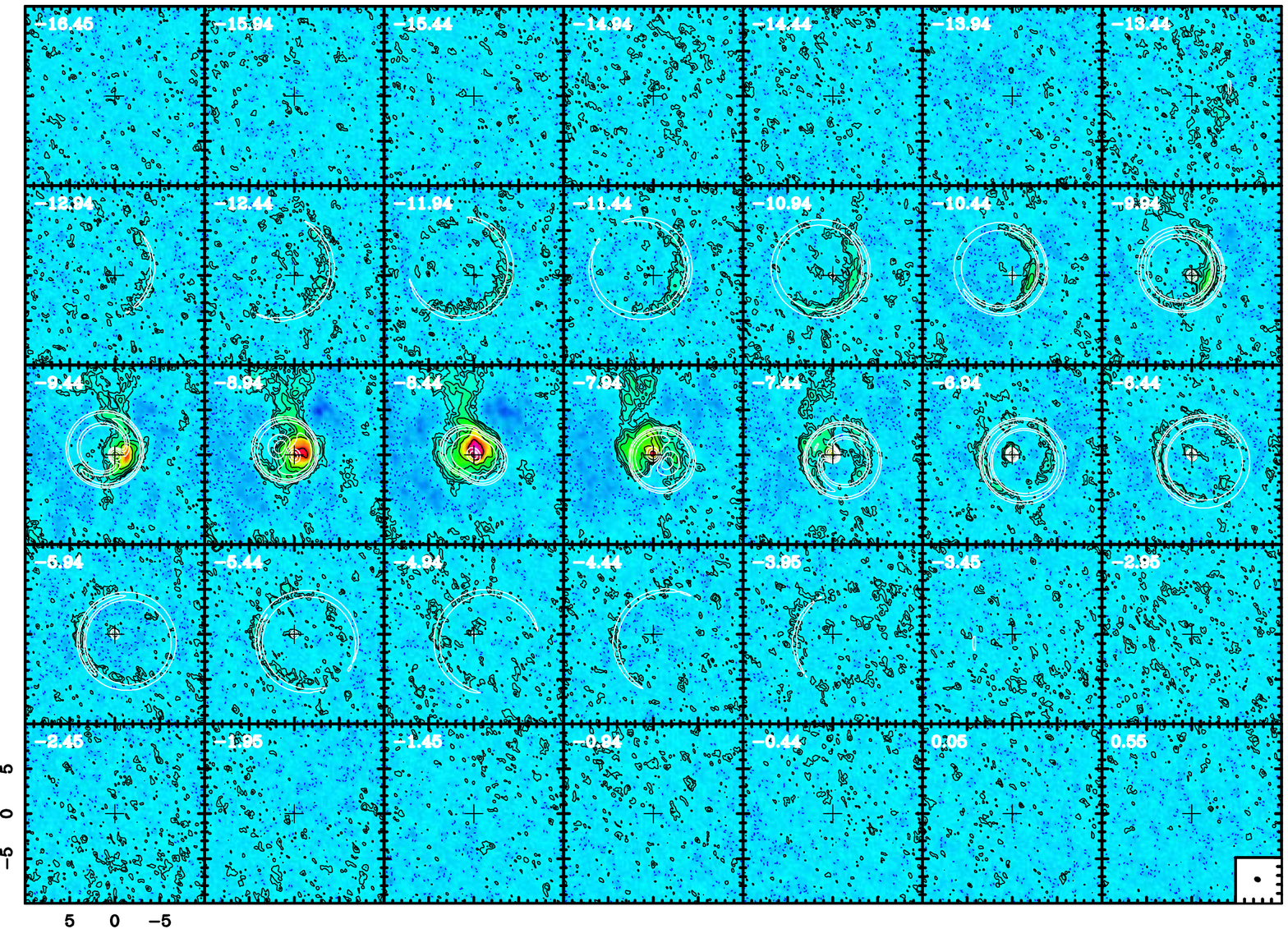}
\caption{\small Synthetic maps predicted by the model of the \trece \dosuno\ line emission for the nebula around 89\,Her, which is shown in white. The maps are superimposed on the observational ones. The contours for both the observational data and model are: $-$5 (observational data only), 5, 10, 20, 40, 80, and 160\,mJy\,beam$^{-1}$.}
    \label{fig:mapas_trece_modelo}  
\end{figure*}

\begin{figure}[h]
\centering
\includegraphics[width=\sz\linewidth]{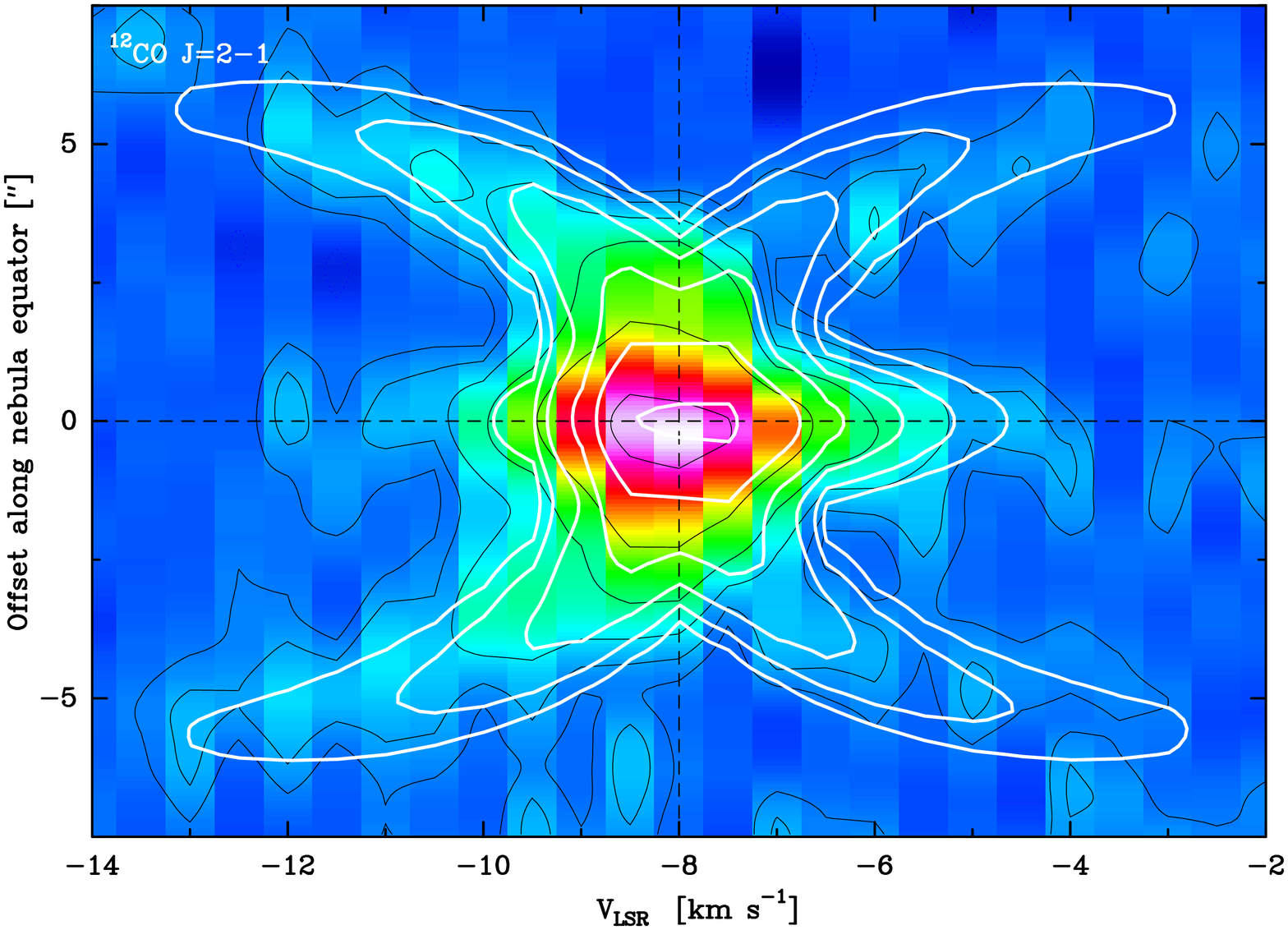}
\caption{\small Synthetic PV diagram predicted by the model of the \doce \dosuno\ line emission in 89\,Her along the equatorial direction (PA = 150\degree), which is shown in white. It is superimposed on the observational one. The contours for both the observational data and model are: $-$40 (observational data only), 40, 80, 160, 320, 640, and 1280\,mJy\,beam$^{-1}$.}
    \label{fig:pv_doce_ec_modelo}  
\end{figure}

\begin{figure}[h]
\centering
\includegraphics[width=\sz\linewidth]{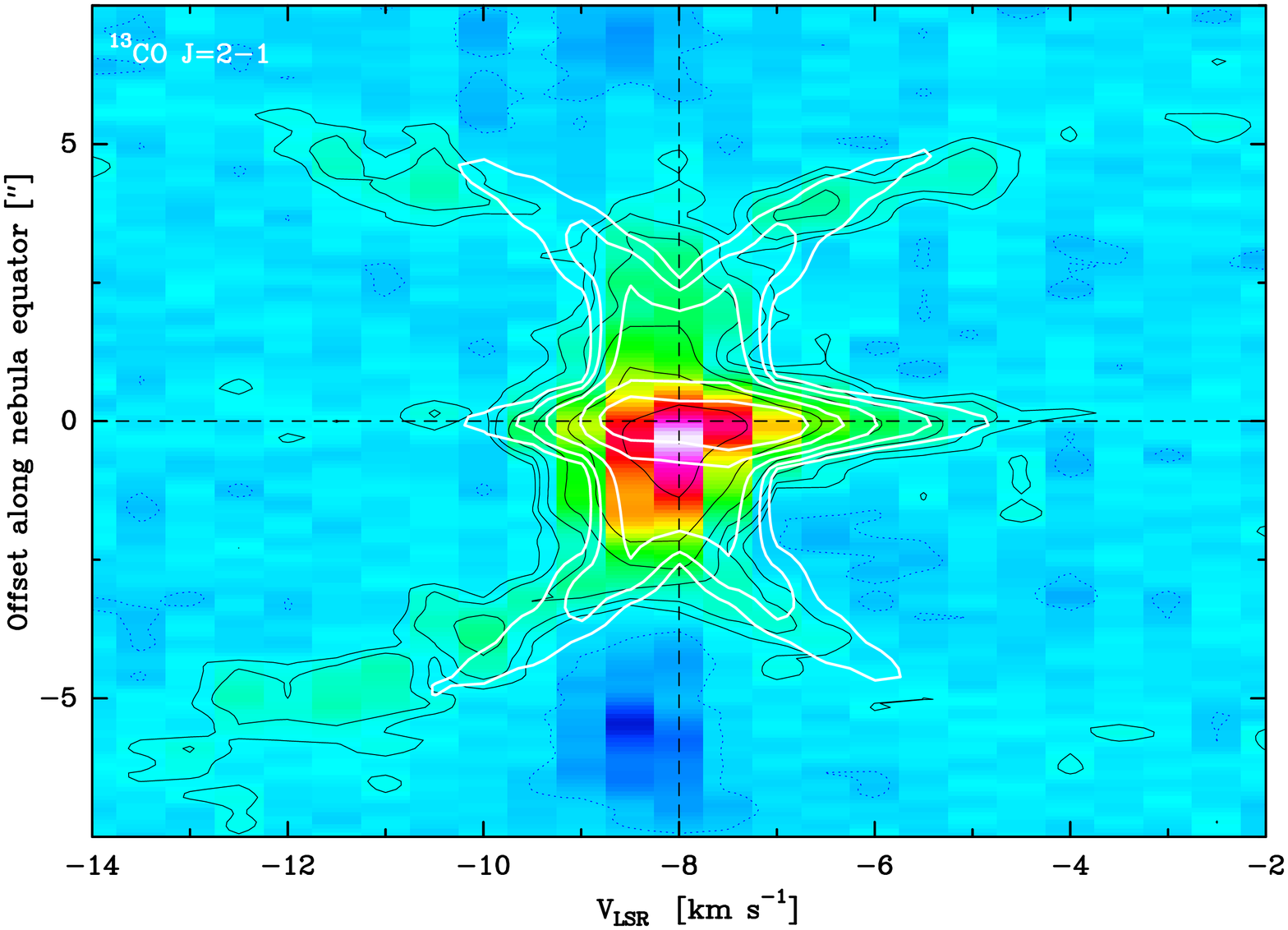}
\caption{\small Synthetic PV diagram predicted by the model of the \doce \dosuno\ line emission in 89\,Her along the equatorial direction (PA = 150\degree), which is shown in white. It is superimposed on the observational one. The contours for both the observational data and model are: $-$5 (observational data only), 5, 10, 20, 40, 80, and 160\,mJy\,beam$^{-1}$.}
    \label{fig:pv_trece_ec_modelo}  
\end{figure}

In this Appendix, we show the synthetic velocity maps of 89\,Her for \doce and \trece \dosuno\ emission for our best-fit model (see \figs\ref{fig:mapas_doce_modelo} and \ref{fig:mapas_trece_modelo}).
The large hourglass-like structure is present in our synthetic velocity maps.
We also show the synthetic PV diagrams along the equatorial direction for \doce and \trece \dosuno\ line emission (see \figs\ref{fig:pv_doce_ec_modelo} and \ref{fig:pv_trece_ec_modelo}).
See \sect\ref{modelo} for further details.

\end{document}